\documentclass[lettersize,journal]{IEEEtran}
\usepackage{amsmath,amsfonts}
\usepackage{algorithmic}
\usepackage{algorithm}
\usepackage{array}
\usepackage[caption=false,font=normalsize,labelfont=sf,textfont=sf]{subfig}
\usepackage{textcomp}
\usepackage{stfloats}
\usepackage{url}
\usepackage{verbatim}
\usepackage{graphicx}
\usepackage{cite}
\usepackage{bm}
\def\BibTeX{{\rm B\kern-.05em{\sc i\kern-.025em b}\kern-.08em
    T\kern-.1667em\lower.7ex\hbox{E}\kern-.125emX}}

\begin{document}

\title{Beyond Static Policies: Exploring Dynamic Policy Selection for Single-Thread Performance Optimization}

\author{\IEEEauthorblockN{Yanxin Zhang$^{1}$, Ian McDougall$^{1}$, Junnan Li$^{1}$, Shayne Wadle$^{1}$, Vikas Singh$^{1}$, Karthikeyan Sankaralingam$^{1,2}$}
\IEEEauthorblockA{$^{1}$\textit{Department of Computer Sciences, University of Wisconsin--Madison}, Madison, WI, USA \\
\{yanxin, imcdouga, junnanl, swadle, vsingh, karu\}@cs.wisc.edu}
\IEEEauthorblockA{$^{2}$\textit{NVIDIA}}}



\maketitle

\begin{abstract}
For over a decade, processor design has focused on implementing sophisticated policies for various components of the out-of-order pipeline, including cache replacement and prefetching. The prevailing design philosophy has been to build processors with a single, static selection of policies across these different mechanisms. This paper investigates a fundamental question: do different workloads, or even different execution phases within the same workload, benefit from different policy combinations? We present a comprehensive analysis exploring whether a hypothetical processor capable of dynamically selecting from multiple policies could significantly outperform traditional static-policy processors. Using ChampSim-based simulation across 49 benchmarks segmented into 490 execution phases of 20M instructions each, we evaluate performance across multiple policy combinations for cache replacement and prefetching. Our findings reveal that significant performance headroom exists: the best static policy achieves optimal performance for only 19.18\% of execution phases and incurs a mean IPC loss of 1.54\% compared to an oracle. Moreover, 85 phases (17.35\%), spanning 14 of the 49 applications, exhibit more than 2.5\% IPC loss relative to the oracle. Furthermore, we demonstrate that a processor capable of dynamically switching between two carefully chosen policies can achieve a 13.6$\times$ reduction in mean IPC loss (from 1.54\% to 0.11\%) and match oracle performance 52.65\% of the time. These results suggest that dynamic policy selection represents a promising avenue for unlocking single-thread performance improvements that have become increasingly difficult to achieve.

\end{abstract}

\begin{IEEEkeywords}
processor architecture, dynamic policy selection, cache replacement, prefetching, performance optimization
\end{IEEEkeywords}

\section{Introduction}

The past decade of processor microarchitecture has been characterized by the development and refinement of increasingly sophisticated policies embedded throughout the out-of-order execution pipeline. These policies govern critical architectural decisions including cache replacement strategies and hardware prefetching mechanisms. Each of these components plays a crucial role in determining overall performance, yet the prevailing design philosophy has been to configure processors with a single, static combination of policies that remains fixed throughout program execution.

This design approach raises a fundamental question that has received limited attention in the architecture community: how much performance headroom exists if processors can dynamically select among multiple policies across execution phases?

\IEEEpubidadjcol To calibrate what counts as a meaningful IPC difference, it is useful to examine how close competing state-of-the-art policies already are when compared apples-to-apples across the same benchmark suite. Among the L1D prefetchers we study, Gaze and Berti are close competitors: in our own dataset, Gaze outperforms Berti on 63.67\% of timesteps, with an average speedup of 2.40\% on those timesteps; it underperforms on 36.33\% of timesteps, with an average slowdown of 3.31\%; there are no ties. Even so, these differences remain in the low single digits, indicating that strong state-of-the-art policies often differ by modest margins rather than by large, uniform gaps.

Viewed in that context, the headroom exposed by dynamic policy selection is meaningful. In our dataset, the best static fixed policy (Berti/Entangling/Mockingjay) incurs a mean IPC loss of 1.54\% relative to the oracle. Moreover, the oracle exceeds the static policy by more than 2.5\% in 85 of 490 timesteps (17.35\%), and these opportunities are spread across 14 of the 49 applications. Relative to a naive IP-stride/LRU baseline, the oracle improves mean IPC by 30.09\% and exceeds 2.5\% improvement in 470 timesteps (95.92\%), with a mean improvement of 31.31\% on those timesteps, spanning 48 of the 49 applications. This breadth indicates that the opportunity is not benchmark-specific.

This work provides a comprehensive answer to these questions through extensive simulation-based analysis. We explore whether dynamic policy selection represents a fruitful research direction---a new knob by which we can unlock single-thread performance improvements that have proven extremely difficult to achieve through conventional means.

Our key contributions include:

\begin{itemize}
\item A unified simulation framework integrating multiple state-of-the-art policies for L1 data cache prefetching, L1 instruction cache prefetching, and L2 cache replacement
\item Comprehensive performance analysis across 49 diverse benchmarks, each segmented into 10 execution phases (timesteps) of 20M instructions, totaling 490 distinct execution scenarios
\item Quantification of the performance gap between static policy selection and an oracle with perfect knowledge: the best static policy achieves 19.18\% oracle match rate and incurs a mean IPC loss of 1.54\%, with 85 phases (17.35\%) across 14 applications exhibiting more than 2.5\% IPC loss relative to the oracle
\item Demonstration that a two-policy dynamic selection mechanism can achieve a 13.6$\times$ reduction in mean IPC loss, while increasing oracle match rate from 19.18\% to 52.65\% of timesteps
\item Analysis of policy selection patterns revealing concentration on a small number of highly competitive policies, with 62.86\% of timesteps already within 0.5\% of optimal using the best static choice
\end{itemize}

Our findings demonstrate that significant performance headroom exists for dynamic policy selection mechanisms. Even the best static policy is optimal for only 19.18\% of execution phases, incurring measurable IPC loss across diverse workloads and execution phases. In 85 timesteps (17.35\%) across 14 applications, the best static policy incurs more than 2.5\% IPC loss relative to the oracle. A simple two-policy dynamic approach captures much of this headroom, achieving a mean loss of just 0.11\% and keeping 96.53\% of timesteps within 0.5\% of optimal. Designing an efficient runtime mechanism to select among policies is an important direction for future work and is beyond the scope of this paper.

\section{Methodology}

\subsection{Simulation Infrastructure}

Our analysis is built upon ChampSim\footnote{\url{https://github.com/ChampSim/ChampSim}}, a widely-used trace-driven microarchitecture simulator that has been extensively validated against real hardware~\cite{gober2022championship}. ChampSim provides a flexible framework for implementing and evaluating various architectural policies while maintaining simulation speed sufficient for large-scale studies.

We extended ChampSim to implement multiple state-of-the-art policies within a unified codebase. This unified implementation is critical for ensuring fair comparison, as it eliminates variability that could arise from different simulator configurations, warmup periods, or measurement methodologies. All policies were implemented with careful attention to published specifications and validated against reported results where possible.

\subsection{Policy Implementations}

Our study evaluates policies across three key microarchitectural components. 
For L1 data cache prefetching, we implement Berti~\cite{navarro2022berti} and Gaze~\cite{chen2025gaze}. 
For L1 instruction cache prefetching, we evaluate Entangling~\cite{ros2021cost} and BARCA~\cite{gratz2020barca}. 
For L2 cache replacement, we implement Mockingjay~\cite{shah2022effective} and PACIPV~\cite{mostofi2025light}.
\subsection{Benchmark Suite}

We selected 49 diverse benchmarks representing a wide range of application domains and execution characteristics. These benchmarks were chosen to provide comprehensive coverage of different computational patterns, memory behaviors, and control flow characteristics.

\subsection{Temporal Segmentation}

A key aspect of our methodology is the temporal segmentation of each benchmark into discrete execution phases. Each benchmark trace was divided into chunks of 20,000,000 instructions, which we refer to as \textit{timesteps}. This granularity was chosen based on several considerations:

\begin{itemize}
\item It is fine enough to capture phase behavior within applications
\item It is coarse enough to amortize any overhead of policy switching
\item It provides sufficient instructions for stable IPC measurements
\end{itemize}

With 49 benchmarks and 10 timesteps per benchmark, our complete evaluation spans 490 distinct execution phases, providing a comprehensive dataset for analysis.

Compared with non-checkpointed runs, checkpointing introduces a history-truncation bias because each chunk begins without the full microarchitectural state accumulated in prior execution; in our data, the mean absolute IPC gap drops from about 17.3\% with 2M-sized chunks to about 4.0\% with 20M-sized chunks, suggesting that this bias becomes increasingly negligible as chunk length grows.

\subsection{Performance Metrics}

For each combination of benchmark, timestep, and policy configuration, we measure Instructions Per Cycle (IPC) as the primary performance metric. 
All IPC values are collected across the entire execution timeline for 
every application and policy configuration under study.

From these raw IPC measurements, we derive several analysis metrics:

\textbf{IPC Loss:} For each timestep, we compute the IPC loss of a static policy relative to the oracle (the best-performing policy at that timestep):
\begin{equation}
\text{Loss} = \frac{\text{IPC}_{\text{oracle}} - \text{IPC}_{\text{policy}}}{\text{IPC}_{\text{oracle}}} \times 100\%
\end{equation}

\textbf{Oracle Performance:} The oracle represents an idealized processor with perfect knowledge of which policy will perform best at each timestep. While not practically realizable, it provides an upper bound on the performance achievable through dynamic policy selection.

\textbf{Baseline Performance:} We define a baseline processor configuration and report its performance under a fixed static policy selection; at minimum, this baseline is specified by issue width, ROB size, and L1/L2 cache sizes. In addition to comparing against the strongest static policy among the eight evaluated combinations, we also use a naive baseline that combines IP-stride prefetching with LRU replacement.

\textbf{Best Static Policy:} The best static policy is the fixed policy combination with the highest average IPC across all 490 timesteps.

\section{Results}

\subsection{Single Policy Analysis: Per-Application Variability}

\begin{figure}[tb]
\centering
\includegraphics[width=0.48\textwidth]{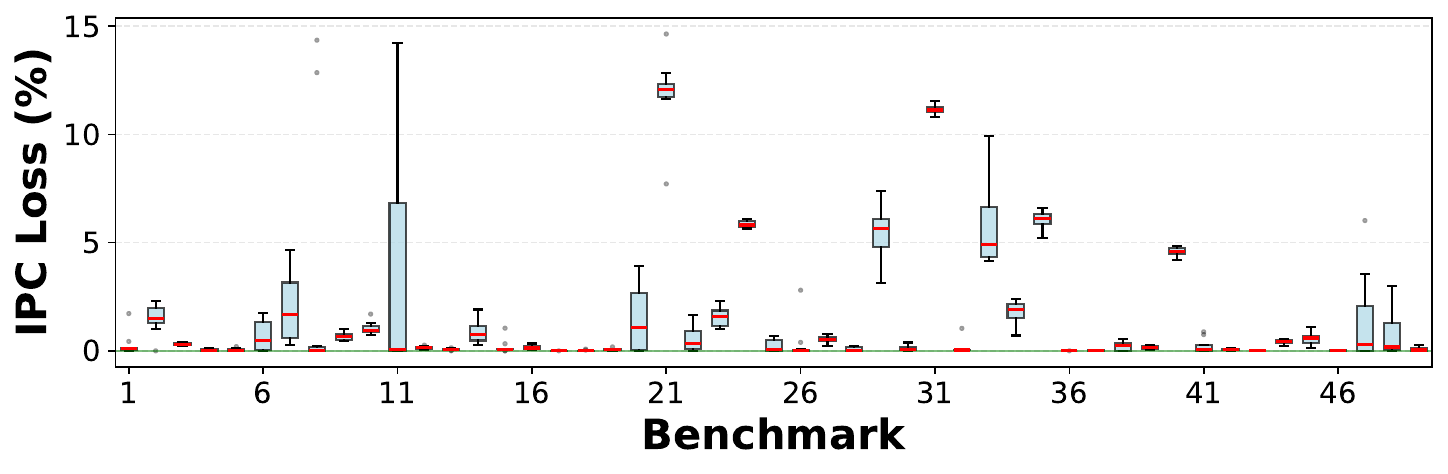}
\caption{Box plot showing the distribution of IPC loss across timesteps for each benchmark when using a single static policy (Berti/Entangling/Mockingjay). Each box represents the distribution across 10 timesteps for one benchmark. Wide distributions indicate high temporal variability, suggesting that different execution phases would benefit from different policies.}
\label{fig:loss_dist_single}
\end{figure}

We first analyze how policy effectiveness varies across execution
phases within individual applications. 
Figure~\ref{fig:loss_dist_single} shows the IPC loss distribution of the
best-by-mean static policy (Berti/Entangling/Mockingjay) across all
benchmarks.

The results reveal substantial temporal variability: even with the best-by-mean static policy, performance varies widely across execution phases, indicating strong potential benefit from dynamic policy selection. Across all 490 phases, IPC loss exceeds 10\% in 23 phases (4.69\%), exceeds 5\% in 58 phases (11.84\%), and exceeds 1\% in 139 phases (28.37\%).

\subsection{Global Static Policy Performance and Policy Comparison}

\begin{figure}[tb]
\centering
\includegraphics[width=0.48\textwidth]{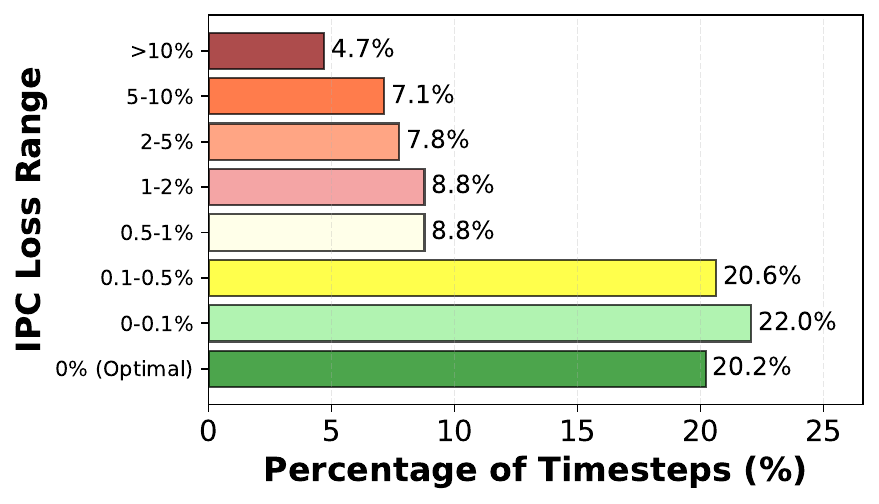}
\caption{Global distribution of IPC loss for a representative static policy compared to the oracle across all 490 timesteps. Each data point corresponds to one 20M-instruction execution phase.}
\label{fig:static_vs_oracle_single}
\end{figure}

To quantify overall performance headroom, we aggregate all 490 timesteps
(49 benchmarks $\times$ 10 timesteps) and treat each timestep as an
independent data point. Figure~\ref{fig:static_vs_oracle_single} shows the
IPC loss distribution of a representative static policy relative to the
oracle.

Even the best-performing static configuration by mean IPC, Berti/Entangling/Mockingjay, achieves a 19.18\% oracle match rate and incurs a mean IPC loss of 1.54\%. Moreover, 28.37\% of timesteps incur more than 1\% IPC loss relative to the oracle, indicating substantial performance headroom for dynamic policy selection.

\begin{figure}[tb]
\centering
\includegraphics[width=0.48\textwidth]{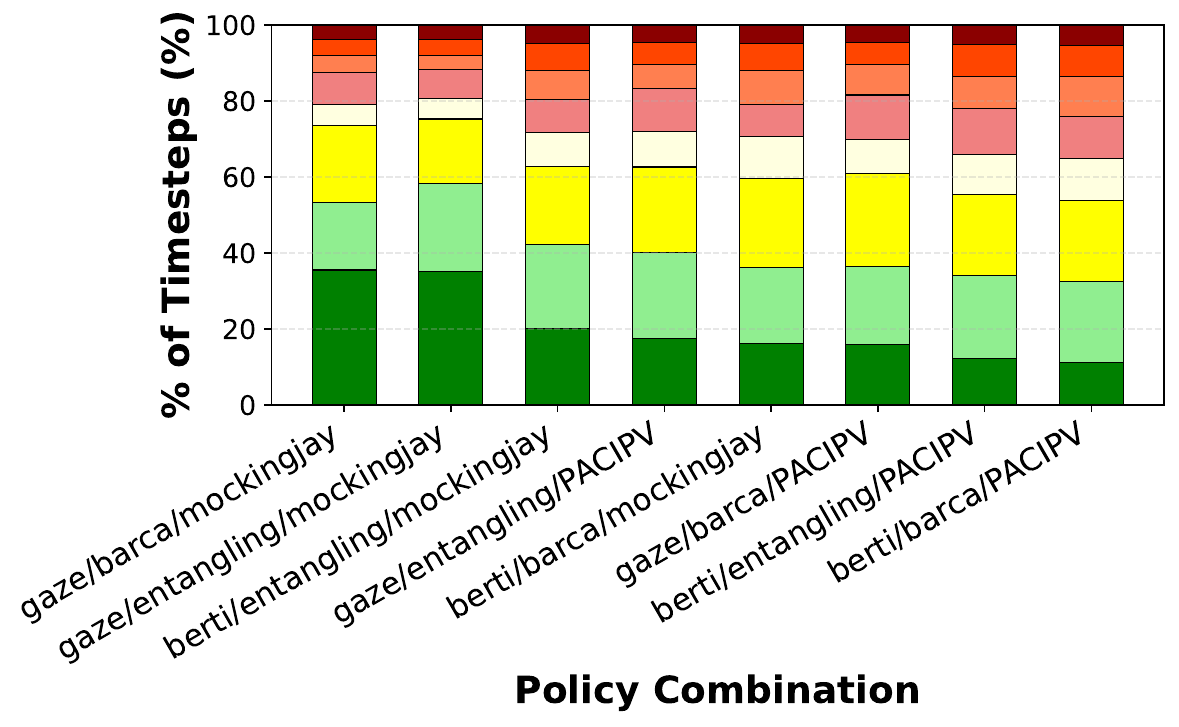}
\caption{Comparison of IPC loss distributions across all evaluated static policy combinations. Each distribution is measured relative to the oracle across all 490 timesteps. The segments, from bottom-to-top, represent the 0\%, 0-0.1\%, 0.1-0.5\%, 0.5-1\%, 1-2\%, 2-5\%, 5-10\%, and $>$10\% IPC loss ranges, respectively. }
\label{fig:all_policies}
\end{figure}

To determine whether this headroom is specific to one configuration,
Figure~\ref{fig:all_policies} compares IPC loss distributions across
all evaluated static policy combinations.

No single static configuration dominates across all timesteps, and
policy effectiveness varies substantially across execution phases,
highlighting the structural limitations of static policy selection.

\subsection{Policy Optimality and Two-Policy Dynamic Selection}

\begin{figure}[tb]
\centering
\includegraphics[width=0.48\textwidth]{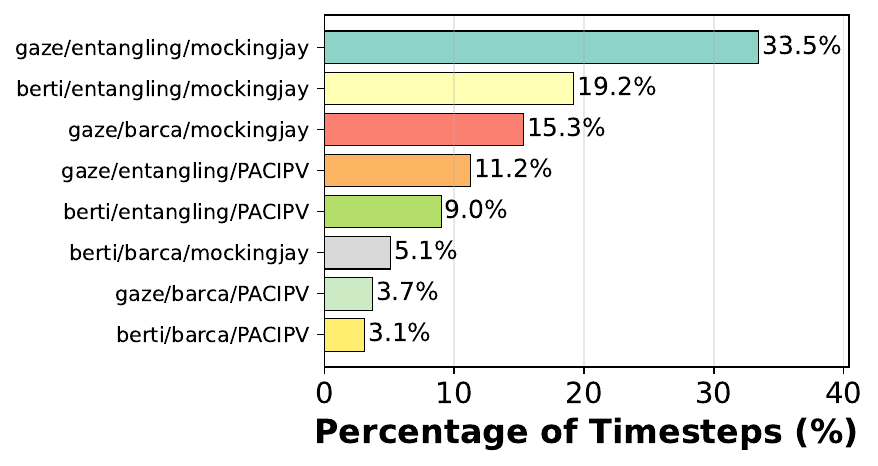}
\caption{Frequency with which each policy configuration is optimal across all 490 timesteps.}
\label{fig:policy_freq}
\end{figure}

To examine whether oracle optimality is concentrated on a small number
of policies, we analyze how frequently each static configuration
achieves the highest IPC at each timestep.
Figure~\ref{fig:policy_freq} shows the optimality frequency across
all 490 timesteps.

Optimality is concentrated on a small number of dominant policy combinations. The most frequently selected winner is Gaze/Entangling/Mockingjay at 33.47\% of timesteps, followed by Berti/Entangling/Mockingjay at 19.18\%; together, these two configurations cover 52.65\% of timesteps. This does not necessarily identify the best static policy by mean IPC, because a policy that wins more frequently can still have lower average IPC if its losses are larger when it is not optimal. This concentration helps explain why a two-policy mechanism can recover most of the oracle benefit despite the larger underlying policy space.

\begin{figure}[tb]
\centering
\includegraphics[width=0.48\textwidth]{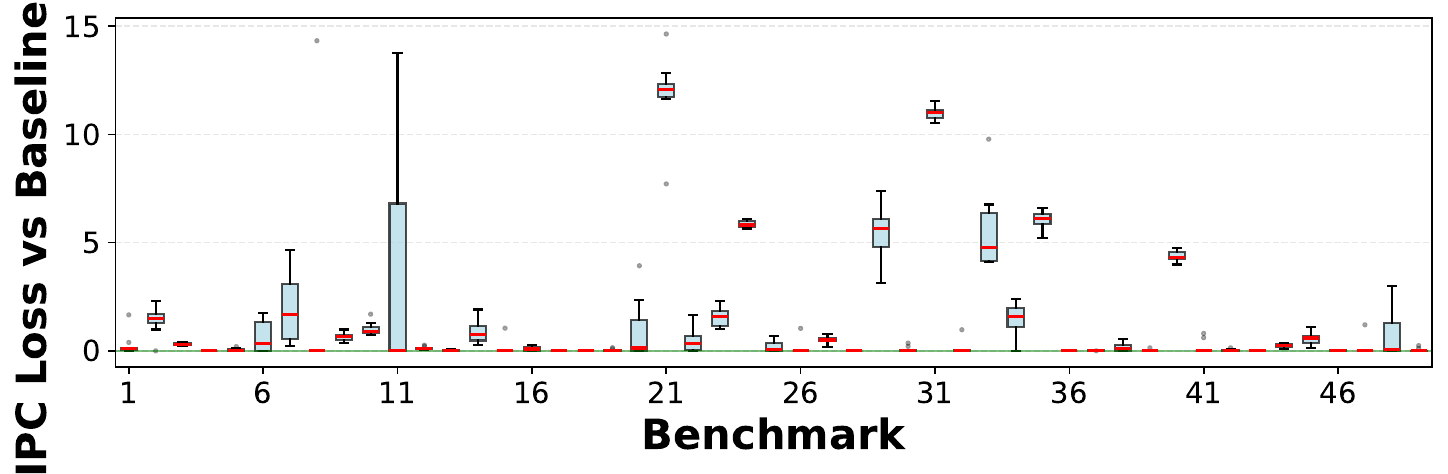}
\caption{IPC loss distribution of the best static baseline (Berti/Entangling/Mockingjay) relative to a two-policy mechanism composed of the two carefully chosen policies.}

\label{fig:two_policy}
\end{figure}

Motivated by this observation, we evaluate whether two carefully chosen policies suffice, selecting the pair by overall loss reduction rather than winner frequency alone. Figure~\ref{fig:two_policy}
compares the best static baseline (Berti/Entangling/Mockingjay) against a two-policy mechanism composed of the two carefully chosen policies.

Compared to the best static configuration (mean IPC loss 1.54\%), the two-policy mechanism reduces mean loss to 0.11\% (13.6$\times$ improvement) and increases the oracle match rate from 19.18\% to 52.65\% of timesteps.

These results show that supporting only two selectable policies
captures most of the oracle's performance advantage while avoiding
the complexity of full dynamic selection.

\subsection{Summary of Key Findings}

Taken together, our results show that static policy selection leaves measurable performance headroom across diverse workloads and execution phases. Because much of this opportunity is concentrated in a small number of highly competitive policies, a simple two-policy mechanism can capture most of the oracle benefit without requiring full dynamic selection.

\section{Related Work}

Prior work has explored adaptive mechanisms across several
microarchitectural components. Adaptive cache replacement
policies such as DRRIP adjust replacement decisions based on
workload behavior, but operate within a single policy framework~\cite{jaleel2010high}. Similarly, industry standard adaptive prefetching techniques are limited to a selection of relatively simplistic policies, although research has demonstrated performance benefits from more complex composite prefetcher options~\cite{alcorta2024characterizing}.

Prior work on phase detection and reconfigurable architectures
has further investigated runtime adaptation of architectural
behavior~\cite{wadle2025sahm}. In contrast, our work quantifies the performance
opportunity of dynamically selecting among multiple
microarchitectural policies across execution phases.

\section{Discussion}

While our results quantify substantial performance headroom, practical
realization of dynamic policy selection requires addressing switching
overhead and phase detection mechanisms.

Our analysis shows that optimality is concentrated on a small number of strong candidate policies, suggesting that a simple two-way selection mechanism with carefully chosen policies can capture most of the oracle’s benefit without supporting full dynamic selection. We do not claim this to be universal; depending on
the policy space considered, more than two policies may be required.

These observations indicate that dynamic policy selection represents a
practical and architecturally tractable opportunity for improving
single-thread performance.
\section{Conclusion}

We quantify the performance headroom left by static microarchitectural policy selection across 49 benchmarks and 490 execution phases. Our results show that optimal policy choice varies across phases and that no single static configuration dominates.

Because optimality is concentrated on a small number of strong candidate policies, a simple two-policy dynamic mechanism with carefully chosen policies captures much of the oracle advantage, reducing mean IPC loss from 1.54\% to 0.11\% and achieving oracle-level performance in 52.65\% of phases.

These findings highlight dynamic policy selection as a practical
direction for improving single-thread performance.

\section*{Acknowledgments}

The authors would like to thank the developers of ChampSim for providing an excellent simulation infrastructure that made this research possible.

\bibliographystyle{IEEEtran}
\bibliography{refs}

@article{gober2022championship,
  title={The championship simulator: Architectural simulation for education and competition},
  author={Gober, Nathan and Chacon, Gino and Wang, Lei and Gratz, Paul V and Jimenez, Daniel A and Teran, Elvira and Pugsley, Seth and Kim, Jinchun},
  journal={arXiv preprint arXiv:2210.14324},
  year={2022}
}

@inproceedings{navarro2022berti,
  title={Berti: an accurate local-delta data prefetcher},
  author={Navarro-Torres, Agust{\'\i}n and Panda, Biswabandan and Alastruey-Bened{\'e}, Jes{\'u}s and Ib{\'a}{\~n}ez, Pablo and Vi{\~n}als-Y{\'u}fera, V{\'\i}ctor and Ros, Alberto},
  booktitle={2022 55th IEEE/ACM International Symposium on Microarchitecture (MICRO)},
  pages={975--991},
  year={2022},
  organization={IEEE}
}

@inproceedings{chen2025gaze,
  title={Gaze into the pattern: characterizing spatial patterns with internal temporal correlations for hardware prefetching},
  author={Chen, Zixiao and Wu, Chentao and Gu, Yunfei and Jia, Ranhao and Li, Jie and Guo, Minyi},
  booktitle={2025 IEEE International Symposium on High Performance Computer Architecture (HPCA)},
  pages={173--187},
  year={2025},
  organization={IEEE}
}

@inproceedings{ros2021cost,
  title={A cost-effective entangling prefetcher for instructions},
  author={Ros, Alberto and Jimborean, Alexandra},
  booktitle={2021 ACM/IEEE 48th Annual International Symposium on Computer Architecture (ISCA)},
  pages={99--111},
  year={2021},
  organization={IEEE}
}

@article{gratz2020barca,
  title={Barca: Branch agnostic region searching algorithm},
  author={Gratz, Daniel A Jim{\'e}nez Paul V and Gober, Gino Chacon Nathan},
  journal={The First Instruction Prefetching Championship},
  year={2020}
}

@inproceedings{shah2022effective,
  title={Effective mimicry of belady’s min policy},
  author={Shah, Ishan and Jain, Akanksha and Lin, Calvin},
  booktitle={2022 IEEE International Symposium on High-Performance Computer Architecture (HPCA)},
  pages={558--572},
  year={2022},
  organization={IEEE}
}

@inproceedings{mostofi2025light,
  title={Light-weight Cache Replacement for Instruction Heavy Workloads},
  author={Mostofi, Saba and Gupta, Setu and Hassani, Ahmad and Tibrewala, Krishnam and Teran, Elvira and Gratz, Paul V and Jim{\'e}nez, Daniel A},
  booktitle={Proceedings of the 52nd Annual International Symposium on Computer Architecture},
  pages={1005--1019},
  year={2025}
}

@article{alcorta2024characterizing,
  title={Characterizing Machine Learning-Based Runtime Prefetcher Selection},
  author={Alcorta, Erika S and Madhav, Mahesh and Afoakwa, Richard and Tetrick, Scott and Yadwadkar, Neeraja J and Gerstlauer, Andreas},
  journal={IEEE Computer Architecture Letters},
  volume={23},
  number={2},
  pages={146--149},
  year={2024},
  publisher={IEEE}
}

@article{wadle2025sahm,
  title={SAHM: State-Aware Heterogeneous Multicore for Single-Thread Performance},
  author={Wadle, Shayne and Sankaralingam, Karthikeyan},
  journal={arXiv preprint arXiv:2509.22405},
  year={2025}
}

@article{jaleel2010high,
  title={High performance cache replacement using re-reference interval prediction (RRIP)},
  author={Jaleel, Aamer and Theobald, Kevin B and Steely Jr, Simon C and Emer, Joel},
  journal={ACM SIGARCH computer architecture news},
  volume={38},
  number={3},
  pages={60--71},
  year={2010},
  publisher={ACM New York, NY, USA}
}

\end{document}